\documentclass[preprint]{aastex}

\slugcomment{To appear in ApJ}

\shorttitle{A Radio Halo in the Cluster 1E0657$-$56} \shortauthors{Liang
et al.}

\begin{document}


\title{A Powerful Radio Halo in the Hottest Known Cluster of
Galaxies 1E0657$-$56}

\author{H. Liang} \affil{Physics Department, University of Bristol,
Tyndall Avenue, Bristol, BS8 1TL, UK} \email{h.liang@bristol.ac.uk}

\author{R. W. Hunstead} \affil{School of Physics, University of
Sydney, NSW 2006, Australia} \email{rwh@physics.usyd.edu.au}

\author{M. Birkinshaw} \affil{Physics Department,
University of Bristol, Tyndall Avenue, Bristol, BS8 1TL, UK}
\email{Mark.Birkinshaw@bristol.ac.uk}

\and

\author{P. Andreani} \affil{Dip. di Astronomia, Universit\`a di
Padova, Vicolo Osservatorio 5, I-35122, Padova, Italy}
\email{andreani@pd.pd.astro.it}

\begin{abstract}
We report the detection of a diffuse radio halo source in the hottest
known cluster of galaxies 1E0657$-$56 (RXJ0658$-$5557). The radio halo has
a morphology similar to the X-ray emission from the hot intracluster
medium.  The presence of a luminous radio halo in such a hot
cluster is further evidence for a steep correlation between the radio halo
power and the X-ray temperature. We favour models for
the origin of radio halo sources involving a direct connection between
the X-ray emitting thermal particles and the radio emitting
relativistic particles.

\end{abstract}


\section{INTRODUCTION}

Diffuse cluster radio sources are found in a few X-ray luminous
clusters of galaxies. They are extended ($\sim 1$ Mpc), have low
surface brightnesses, and exhibit steep spectra ($\alpha \leq -1$,
$S_\nu \propto \nu^{\alpha}$). They cannot be identified with any one
individual galaxy but are associated with the cluster as a
whole. Diffuse cluster radio sources are generally separated into two
classes: halos and relics. Halos are centred on the X-ray
emission (e.g. Coma-C is a proto-type radio halo source; Giovannini et
al. 1993) whereas relics are peripheral and exhibit stronger
polarisation than halos (e.g. A3667 has a large relic; R\"ottgering et
al. 1997). In this paper, we will concentrate on halo sources.

Until recently, systematic surveys for radio halo sources found few
examples, with the total number of known halos being $\sim 5$ (Feretti
\& Giovannini 1996). They are thus considered to be rare, and owing to
their small number remain a poorly understood class of radio sources
even though the first example, Coma-C, was discovered over 20 years
ago. The spectra suggest that halo radio emission arises predominantly
by the synchrotron process. However, the formation of radio halos
remains a puzzle: why do they occur in some clusters and not in
others, and what is the origin of the magnetic field and relativistic
particles?

A number of models have been proposed to explain the formation of
radio halos (e.g. Jaffe 1977; Dennison 1980; Roland 1981). Most of
these early models suggest that ultra-relativistic electrons originate
either as relativistic electrons from cluster radio sources
re-accelerated by in-situ Fermi processes or turbulent galactic wakes,
or as secondary electrons produced by the interaction between
relativistic protons (again from cluster radio galaxies) and thermal
protons.  However, the energetics involved are problematic and the
models could not always fit the observations (e.g., see review by
B\"ohringer 1995). Harris et al. (1980) first suggested that radio
halos are formed in cluster mergers where the merging process creates
the shocks and turbulence necessary for the magnetic field
amplification and high energy particle acceleration. More recently,
Tribble (1993) showed that the energetics involved in a merger are
more than enough to power a radio halo. The halos thus produced are
expected to be transient since the relativistic electrons lose energy
on time scales of $\sim 10^{8}$\,yrs and the time interval between
mergers is of order $\sim 10^{9}$\,yrs. This argument was used to
explain why radio halos are rare.

In this paper, we will describe the properties of the radio halo found
in one of the hottest known clusters 1E0657$-$56, and suggest a new
model for the origin of cluster halos based on the radio and X-ray
properties of all 10 confirmed halos.  \S\ref{s2} describes the
multi-wavelength properties of cluster 1E0657$-$56; \S\ref{s3}
describes the radio observations; \S\ref{s4} discusses the radio
properties of the halo found in 1E0657$-$56; and \S\ref{s5} discusses
the origin of radio halos. Throughout the paper we will use $H_{0}=50$
km\,s$^{-1}$\,Mpc$^{-1}$, $q_{0}=0.5$ and $\Lambda_{0}=0$.
  
\section{THE CLUSTER 1E0657$-$56 \label{s2}}

The cluster 1E0657$-$56 was originally found in the Einstein slew
survey (Tucker et al. 1995), and subsequent optical observations
confirmed it to be a rich cluster at $z\sim 0.296$ with a velocity
dispersion of $1213^{+352}_{-191}$ km\,s$^{-1}$ (Tucker et
al. 1998). It has high X-ray luminosity and was shown to be one of the
hottest known clusters by Tucker et al. (1998).

\subsection{X-ray Properties \label{s:xray}}

The cluster was observed for 25\,ksec by ASCA in 1996 May and with the
ROSAT High-Resolution Imager (HRI) in 1995 for 58 ksec (Tucker et
al. 1998).  Tucker et al. (1998) analysed the ASCA Gas
Scintillation Imaging Spectrometers (GIS) and Solid-state Imaging
Spectrometers (SIS) data
and found the cluster to have a best fit temperature of $kT_{x}\sim
17.4\pm 2.5$\,keV and a bolometric luminosity of $L_{bol} \sim (1.4\pm
0.3) \times 10^{46}$ erg s$^{-1}$.  However, Yaqoob (1999) challenged
these results by re-analysing the ASCA GIS/SIS data and arrived at a
temperature of $\sim 11-12$\,keV, abundance $A\sim 0.2$ solar, and a
neutral hydrogen absorption column density $N_H \sim 15\times
10^{20}$\,cm$^{-2}$, much higher than the Galactic value. Yaqoob found
that the only way to reproduce the high temperature deduced by Tucker
et al. (1998) was to fix the neutral hydrogen column density to the
Galactic value, and concluded that fixing the absorption column
density in such a way leads to an artificially high temperature.

Since then, some ROSAT Position-Sensitive Proportional Counter (PSPC)
data have become publicly available. Since it is known that ASCA SIS
data below 1 keV suffer from inaccurate calibration, we re-estimate
the cluster temperature by fitting to the combined ASCA GIS and
ROSAT PSPC data. The GIS and PSPC data complement each other: the GIS
is more sensitive to the cluster temperature while the PSPC is more
sensitive to the soft X-ray absorption.

We followed the standard ASCA procedure for screening the GIS2 and
GIS3 data as set out in {\it The ABC guide to ASCA data
reduction}\footnote{\url{http://heasarc.gsfc.nasa.gov/docs/asca/abc/abc.html}}. The
spectra were extracted from a circular region of radius $7.25'$
centred on the cluster, after the subtraction of a local background,
extracted from the same frame in areas with no obvious emission. The
spectra were regrouped to a minimum of 50 counts per bin. The {\small
XSPEC}
package\footnote{\url{http://legacy.gsfc.nasa.gov/docs/xanadu/xspec/u\_manual.html}}
(Arnaud 1996) was used to fit the GIS spectra between 0.8 and 10\,keV
by a Raymond-Smith spectrum (Raymond \& Smith 1977), with fractional
solar relative abundances from the table of Feldman (1992) and
photoelectric absorption (Morrison \& McCammon 1983). Temperature,
abundance $A$, absorption (parametrised by $N_{H}$), and the
normalisation were taken as free parameters. The best fit was $kT_{x}
\sim 15.6^{+3.1}_{-2.3}$\,keV, $N_{H}\sim (2.2^{+3.9}_{-2.2})\times
10^{20}$\,cm$^{-2}$, and $A \sim 0.49\pm0.27$ with a reduced $\chi^{2}\sim
1.03/{\rm d.o.f.}$. The errors correspond to 90\% confidence limit. It
is clear that $N_H$ is poorly constrained by the GIS data alone, as
expected from the absence of low-energy data.

We retrieved from the ROSAT archive a 4.7\,ksec-exposure PSPC event file
observed in 1997 February. A spectrum of the X-ray emission from the
cluster gas was extracted from the central $5^{'}$ radius after the
subtraction of discrete sources and background contributions. The
background was estimated from an annulus between radii of 8--10$'$
centred on the cluster. Only the spectrum within the energy range
between 0.1 and 2.0\,keV was used for model fitting.

A combined fit of a Raymond-Smith spectrum to the GIS and PSPC spectra
gave the best fit as follows: $kT_{x}=14.5^{+2.0}_{-1.7}$\,keV,
$N_{H}=(4.2^{+0.6}_{-0.5}) \times 10^{20}$\,cm$^{-2}$, $A =
0.33\pm0.16$ (see Fig.~\ref{ascasp}) with a reduced $\chi^{2}\sim
1.03/{\rm d.o.f.}$.  A direct radio-astronomical measure of the
Galactic neutral hydrogen column density towards 1E0657$-$56 gives
$N_{H} \sim 5.8\times 10^{20}$\,cm$^{-2}$ (E.\ M.\ Arnal, private
communication); Dickey \& Lockman (1990) gives a value of $6.6\times
10^{20}$\,cm$^{-2}$. Recently we have obtained a high resolution
($16^{'}$ beam) measurement of the Galactic neutral hydrogen column
density towards 1E0657$-$56 using the Parkes telescope, giving a
$N_{H} \sim 4.6\times 10^{20}$\,cm$^{-2}$ (C. Br\"uns, private
communication) which is very close to the X-ray fitted value.  Thus it
appears that the temperature is lower than that estimated in Tucker et
al. (1998), although 1E0657$-$56 is still one of the hottest known
clusters.  On the other hand, our best fit temperature is higher than
that deduced by Yaqoob (1999) and our best fit $N_{H}$ is much closer
to the Galactic value than that given by Yaqoob. The difference is
unlikely to be caused by different background subtraction techniques,
since Yaqoob tried a wide range of background subtraction methods and
found that the systematic differences between the various techniques
are less than the statistical errors. We tested our result by fixing
the $N_{H}$ and abundance $A$ at the values given by Yaqoob, and found
a best fit $kT_{x}\sim 11.3\pm 1.0$\,keV, consistent with his result
but with a high reduced $\chi^{2}$ of
1.56/d.o.f.. Figure~\ref{ascasp2} shows that Yaqoob's fit (to the ASCA
data alone) is inconsistent with the PSPC data which extend to lower
energies ($\sim 0.1$\,keV) than the SIS ($\sim 0.5$\,keV), and are
therefore more sensitive to soft X-ray absorption.

\placefigure{ascasp} \placefigure{ascasp2}

The HRI image of the cluster shows  two clearly separated
clumps (see Figure~\ref{hrir}).  Andreani et al. (1999) analysed the
spatial distribution of the X-ray emission using the HRI data and
found that the X-ray surface brightness can be fitted with two
spherically symmetric $\beta$-models (Cavaliere \& Fusco-Femiano 1976)
with gas density distributions given by
\begin{equation}
n_{e}(\theta) = n_{e,0}
[1+(\frac{\theta}{\theta_{c}})^{2}]^{-3\beta/2},
\end{equation}
where $\beta=0.7, 0.49$, $\theta_{c}=1.23^{'},0.26^{'}$ and
$n_{e,0}=0.0063, 0.015$ cm$^{-3}$ for the eastern and western clumps
respectively.

\subsection{Sunyaev-Zel'dovich Effect\label{sze}}

The Sunyaev-Zel'dovich effect (SZ effect) is the distortion of the
blackbody spectrum of the Cosmic Microwave Background (CMB) due to
inverse Compton scattering of CMB photons by free electrons in a
plasma such as an intracluster medium (Sunyaev \& Zel'dovich 1972).
The cluster 1E0657-56 was selected as a candidate for the detection of
the SZ effect in the 1994--95 Swedish ESO Submillimeter Telescope
(SEST) campaign (Andreani et al. 1999).  The SEST observations show a
$\sim 4\sigma$ detection of the SZ effect at 1.2\,mm ($\sim$ 150\,GHz)
and a $\sim 3\sigma$ detection at 2\,mm ($\sim$ 250\,GHz) (Andreani et
al. 1999).  By combining the SEST results at 2mm with the X-ray
surface brightness and temperature results (assuming an isothermal
$kT_{x}\sim 17$\,keV), Andreani et al. (1999) deduced a Hubble
constant of $H_{0}=53^{+38}_{-28}$\,km\,s$^{-1}$\,Mpc$^{-1}$.

Shortly after the SEST observations, we obtained data with the
Australia Telescope Compact Array (ATCA) at 8.8 GHz to confirm the
SEST detection of the SZ effect. The ATCA observations were conducted
using the special 210-m array which proved to be an excellent
configuration for detecting low surface brightness diffuse
emissions. The 4.9 and 8.8\,GHz observations show complex radio source
structures, including at least two extended, diffuse sources and an
unusual strongly polarised ($P\sim 60\%$ at 8.8\,GHz) steep spectrum
($\alpha \sim -1.9$) source (Liang et al., 2000, in preparation). The
diffuse radio sources make the detection of the SZ effect difficult
since the SZ effect is also extended and the straightforward
point-source subtraction technique fails.  However, since the radio
sources in the cluster are interesting in their own right, we obtained
further radio data. It was these later observations that led to the
detection of a radio halo source.

\section{RADIO OBSERVATIONS \label{s3}}
\subsection{ATCA Data}

The ATCA has five 22-m antennas on a 3-km east-west rail-track and a
sixth antenna 3\,km from the western end of the track, giving
baselines up to 6\,km.  Simultaneous observations were made in two
frequency bands each of bandwidth 128\,MHz divided into 32 frequency
channels. The cluster 1E0657$-$56 was observed at 1.3, 2.4, 4.9, 5.9,
and 8.8\,GHz in several antenna configurations, so that similar uv
coverage was obtained at all frequencies. Table~\ref{t:obs} gives a
summary of all the radio observations.  The primary flux calibrator
PKS\,B1934$-$638 was observed at least once a day and the phase
calibrator PKS\,B0742$-$56 was observed every $\sim 20$\,min. The data
were calibrated with the {\small MIRIAD} package\footnote{\url{http://www.atnf.csiro.au/computing/software/miriad/userhtml.html}} (Sault et al. 1995).

A radio halo source was clearly detected in maps with resolution $\sim
60^{''}$ in all frequencies (see Fig.~\ref{allfreq}).  A high
resolution 1.3\,GHz image of the cluster field is shown in
Fig.~\ref{21cm}a, where we see the halo source at the cluster centre,
a possible relic source to the east, and a number of tailed sources on
the periphery.

\subsection{MOST Data}

The Molonglo Observatory Synthesis Telescope (MOST) is an east-west
synthesis array comprising two collinear cylindrical paraboloids each
11.6~m wide by 778~m long, separated by a 15~m gap (Mills 1981,
Robertson 1991).  The telescope operates at 843~MHz, with a detection
bandwidth of 3.25~MHz, and forms a comb of 64 real-time fan beams
spaced by $22''$, which are interlaced to a spacing of $11''$.  The
synthesized beam-width is $43'' \times 43'' {\rm cosec} |\delta|$ FWHM
(RA $\times$ Dec).  The observations of 1E0657$-$56 were made as
part of the Sydney University Molonglo Sky Survey (SUMSS, Bock et al.\
1999), in which the pointing of the beam set was time shared among
seven adjoining positions to give a field size of $164' \times 164'$
cosec$|\delta|$.  The background noise in this 12-hour image
was 1.1 mJy rms and the reduction followed the standard survey
pipeline.  For comparison with the ATCA images, the MOST image in
Fig.~\ref{allfreq}a was convolved out to a $60''$ circular beam.

\placefigure{allfreq}

\section{A RADIO HALO IN 1E0657$-$56 \label{s4}}

\subsection{Subtraction of Discrete Sources \label{s:sub}}

In order to obtain a high-quality image of the diffuse radio halo and
estimate its total integrated flux density, it is necessary to
separate the emission from discrete sources embedded in the halo.

One way of estimating the flux density of discrete sources is to
make a high resolution image using only the long baseline data ($>
5000\lambda$) that is unlikely to contain any signal from the diffuse
halo emission. This image, which has a synthesized beam FWHM $\sim
6^{''}$, is then deconvolved using the {\small CLEAN} algorithm.
Fig.~\ref{21cm} shows the image made from all the 1.3\,GHz data and
the image from the long-baseline data with only sources within the
halo region restored.  The clean components of discrete sources
embedded in the halo are then subtracted from the visibility (or uv)
data set. These sources, marked in Fig.~\ref{21cm}b, are unresolved
with the exception of two sources (A,C) which were slightly
extended. The flux densities of these discrete sources at various
frequencies are listed in Table~\ref{ptsrc}. A radio image of the halo
at 1.3 GHz, with these embedded discrete sources subtracted, is shown
in Fig.~\ref{hrir}a \&~\ref{hrir}b.

\placefigure{21cm} \placefigure{hrir}

\subsection{Halo Total Flux Density \label{s:totf}}

Estimation of the total flux density of a diffuse source is rather
difficult, even after solving the problems of separating discrete
sources from the diffuse emission (\S\ref{s:sub}). The difficulty is two-fold:
firstly, we need to define the angular size of the emission, which
depends on the noise level in the image; and secondly, we need to know
if we have sampled a large enough angular scale (i.e. short enough
baselines) to include all the flux density.  In the following
paragraphs we describe two independent methods of estimating the
integrated flux density as a function of the area of integration.

First, we make a low resolution image ($\sim 60^{''}$ beam) using the
uv data-set that has the discrete sources already subtracted, but
taking only the short-baseline data ($<3600\lambda$). This ensures
that the data used for the estimation of the halo flux density are
independent of the long-baseline data used for the subtraction of
discrete sources. The integrated flux density of the halo estimated
from such an image is plotted as a function of the area of integration
in Fig.~\ref{totflux} (open circles). While the low resolution of the
image ensures relatively high brightness sensitivity, the area of
integration is limited because of blending with extended sources
near the halo, a problem made worse by the large beam size (see
Fig.~\ref{21cm}a).

We now examine an alternative method of estimating the total halo flux
density which also serves as a check on the method above. We integrate
the signal within increasing regions of the halo on the high
resolution image (Fig.~\ref{21cm}a) which uses all the data and has
not undergone source subtraction. The flux densities of the sources
embedded in these regions are estimated from the image in
Fig.~\ref{21cm}b (see Table~\ref{ptsrc}) and subtracted from the
integrated flux density, giving the open triangles shown in
Fig.~\ref{totflux}. By careful exclusion of extended sources in or
near the halo, the halo can be integrated over a larger area. In the
first method (open circles in Fig.~\ref{totflux}) the largest area of
integration corresponds to region `1' marked in Fig.~\ref{21cm}b.  The
largest area used in the second method (open triangles) corresponds to
the full extent of the X-ray emission less the regions of other
extended sources. The smallest region in Fig.~\ref{totflux}
corresponds to region `2' in Fig.~\ref{21cm}b.

Fig.~\ref{totflux} shows that the two methods agree and that the halo
emission extends over a region at least $3.5$\,Mpc$^{2}$ in area with
a total flux density of $78\pm5$ mJy at 1.3 GHz, which corresponds to
a rest frame 1.4\,GHz power of $(4.3\pm0.3)\times
10^{25}$\,W\,Hz$^{-1}$. This power is likely to be a lower limit since
there are negative contours around the halo in the 1.3\,GHz image
(Fig.~\ref{allfreq}b), which we interpret as evidence of missing short
spacings.  Further ATCA observations in the 210-m configuration are
needed to study better the outer parts of the halo.  With a power
$P_{1.4} \sim 4.3\times 10^{25}$\,W\,Hz$^{-1}$, and a largest linear
extent $\sim 2$\,Mpc, the radio halo source in 1E0657$-$56 is the
strongest and largest known.

\placefigure{totflux}

\subsection{Correction for the SZ Effect}

As mentioned in \S\ref{sze}, this cluster shows a relatively strong SZ
effect which produces a decrement at centimetre wavelengths and is
also cluster-wide and diffuse. It is thus difficult to separate the SZ
effect from the halo emission using the radio data alone. However, the
SZ effect is strong enough at 4.9, 5.9 \& 8.8\,GHz to cause a
significant underestimate of the radio halo flux density and an
apparent steepening of the spectrum. 

To obtain a reliable radio spectrum for the halo, we need to correct
for the SZ effect at each frequency. We simulate the SZ effect at
4.9 to 8.8\,GHz, using the X-ray data and the SEST detection at
2\,mm. Andreani et al. (1999) found that the SZ effect detected at
2\,mm was consistent with a model of the X-ray surface brightness
expressed by two $\beta$-models (corresponding to the two X-ray
clumps) assuming an isothermal gas temperature of $\sim 17$\,keV, if
$H_{0}\sim 50$\,km\,s$^{-1}$\,Mpc$^{-1}$. For consistency with the
SEST results, we use this gas temperature only for these calculations;
in the rest of the paper, we use the newly-derived gas temperature
given in \S\ref{s:xray}. Since the SEST results provide no structure
information for the gas, we have to use the X-ray derived gas
parameters, i.e. $n_{e,0}, \theta_{c}, \beta$ from Andreani et
al. (1999; or see \S\ref{s:xray}), to simulate an SZ effect image at
each frequency. This simulated SZ effect image is then weighted by the
primary beam of the ATCA, Fourier transformed, and subtracted from the
uv data-set at that frequency. The image obtained from the resulting
uv data-set contains only the radio emission from the halo and can be
analysed as in \S\ref{s:totf} to get the halo flux densities. Fig.~\ref{halosp}
shows the halo flux density in a specific region at various
frequencies before and after the correction for the SZ effect,
indicating a substantial difference at high frequencies.

\subsection{Radio Spectrum}

To obtain the spectral index of the halo, we took data from the inner
uv-plane (baselines $< 3600\lambda$) for each available frequency. The
observations were planned such that roughly the same region in the
uv-plane was sampled at the various frequencies.  The data were
tapered such that the synthesized beam FWHM was $\sim 60^{''}$ at each
frequency. The images are shown in Fig.~\ref{allfreq}.  To form a
spectrum, the same area of integration was used at each frequency and
methods in \S\ref{s:sub} \& \S\ref{s:totf} were used to correct for
small-angular-size radio sources and the SZ effect. The areas were
selected to include only regions of obvious emission at all
frequencies. The radio spectrum between 1.3 and 8.8\,GHz for region
`2' marked in Fig.~\ref{21cm}b is shown in Fig.~\ref{halosp} (dots).
Determination of the spectral index from region `2' is complicated
since it includes three point sources (B, D \& E) and an extended
source (A, see Fig.~\ref{21cm}b). The three point sources are detected
only in the 1.3\,GHz image.  We can obtain the flux density of the
extended source (A) from images like that of Fig~\ref{21cm}b at each
frequency and subtract them from the total integrated flux.  The
resulting spectral index of region `2' is $\sim -1.2$ between 1.3 and
4.9\,GHz, and $\sim -1.3$ between 2.4 and 8.8\,GHz, indicating no
significant steepening of the spectral index between 1.3 and 8.8\,GHz.

We use the data from MOST to extend the spectrum to lower
frequencies. The MOST images were produced by a real-time FFT and are
thus fixed in resolution, excluding the possibility of subtracting the
discrete sources embedded in the halo. We can only form a spectrum for
the central part (region `3') of the halo that is devoid of discrete
sources. The ATCA data at various frequencies were tapered such that
the final resolution would match the MOST resolution of $51^{''}\times
43^{''}$ (${\rm PA} = 0^{\circ}$).  The spectrum of region `3' is
shown by the triangles in Fig.~\ref{halosp}. The spectrum can be
fitted with a power law with a spectral index of $-1.4\pm 0.1$ using a
least squares fit. In comparison, the average spectral index for the
larger region (region `2') is $\sim -1.3\pm 0.1$.  There is no
evidence of any steepening of the spectral index in the outer regions
compared with the central region (unlike what is observed in the
Coma cluster Giovannini et al. 1993). The areas of the regions `3'
and `2' marked in Fig.~\ref{21cm}b are 0.14\,Mpc$^{2}$ and
0.78\,Mpc$^{2}$ respectively.

\placefigure{halosp}

\subsection{Polarisation}
The ATCA measures all components of linear polarisation, and takes
account of the polarisation leakage terms in the calibration of the
data by the {\small MIRIAD} package.

Polarised emission was not detected for the radio halo source. We
obtain an upper limit for linearly polarised emission at 1.3\,GHz of
20\%, 6.5\%, and 1.4\% at resolutions of $10^{''}, 20^{''}$, and
$60^{''}$ (corresponding to linear sizes of 55, 110, and 327\,kpc)
respectively. The upper limit was calculated as
$\sqrt{(Q^{2}+U^{2})}/I$, where $I$ was the Stokes $I$ peak halo flux
density and $Q$ and $U$ were the $3\sigma$ of the Stokes $Q$ and $U$
maps respectively. The polarisation upper limits obtained at higher
frequencies were poorer, because of the steepness of the halo
spectral index and consequent decrease in signal to noise of the data.
So far polarisation has not been detected in any halo source.

\subsection{Pressure \& Energy}
The total energy density, or pressure, is minimum when the energy
density of the relativistic particles is close to the energy density
of the magnetic field (Burbidge 1958). There is no strong physical
justification for the particle and field energies to be in
equipartition, but it has been conjectured that they maybe close to
equipartition (Longair 1997).  If we make this assumption, then we can
calculate the minimum pressure of radiating electrons and magnetic
fields in the halo source and compare it to the thermal pressure from
the X-ray gas. We estimate the minimum pressure by following the
procedure set out in Pacholczyk (1970), assuming a constant spectral
index of $\sim -1.3$ between the cutoff frequencies of 10\,MHz and
100\,GHz, a emission volume filling factor $\phi$ of 1 and a ratio $k$
between the energy in relativistic protons and electrons of 1. The
minimum pressure of the radio plasma is a factor $\sim 10^{4}$ smaller
than the thermal pressure. Since $P_{min} \propto \phi^{-4/7}
(1+k)^{4/7}$, we need $k>10^{6}$ for $P_{min}$ to match the thermal
pressure which seems unlikely.  Similar results have been found in
other radio halos (e.g. Feretti et al. 1997). Since the radio plasma
interpenetrates the thermal gas, the radio and the thermal plasma each
contributes a partial pressure to the total plasma pressure which
appears to be dominated by the thermal pressure.

The total energy of the thermal plasma is $\sim 10^{63}$\,ergs, and
the energy in the relativistic plasma is $\sim 10^{60}$\,ergs under
conditions of equipartition (i.e. $B \sim 1\mu$G).

\subsection{Morphological Structure} 
The radio halo is similar in extent and overall appearance to the
cluster X-ray emission (see Fig.~\ref{hrir}a).
While Fig.~\ref{hrir}a shows that the radio emission is enhanced at the
main peak of the X-ray emission, Fig.~\ref{hrir}b shows that the radio
emission is also enhanced at the densest part of the optical galaxy
distribution. On the whole the radio emission follows that of
the optical galaxy distribution more closely than the X-ray
emission. This was also found in the Coma cluster (Kim et
al. 1990). We will discuss this point further in \S\ref{s53}.

The main concentration of galaxies is displaced from the
dominant (eastern) peak of the X-ray emission (see
Fig.~\ref{hrir}c). Both the X-ray gas and the galaxy distribution
suggest that the cluster is undergoing a merging process. Since the
X-ray clumps are well separated, the merger appears to be in a
relatively early stage. Since galaxies are more or less collisionless,
but gas is collisional, a merger between two subclusters tends to
allow the galaxies to stream past one another, while the gas tends to
coalesce quickly. Fig.~\ref{hrir}c shows that the galaxy clumps in
1E0657$-$56 are further apart than the gas clumps, indicating that the
galaxies have crossed each other at least once. The projected merging
axis appears to be close to the RA-direction. Shocks produced during
the merger are a plausible source for the energy of the relativistic
electrons responsible for the radio halo emission.

\section{THE ORIGIN OF RADIO HALOS \label{s5}}
\subsection{Are Radio Halos Intrinsically Rare?}
A number of surveys have been conducted to search for radio halos. The
earliest were conducted at Green Bank at 610 MHz (Jaffe \& Rudnick
1979), at metre wavelengths 50--120 MHz (Cane {\em et al.} 1981) and at
Arecibo at 430 MHz (Hanisch 1982), but yielded few examples. Most of
the surveys selected either nearby Abell clusters (Hanisch 1982), or
clusters with known X-ray emission or radio sources. More recently,
Lacy {\em et al.} (1993) imaged a sample of radio sources from the 8C
38\,MHz survey (within $3.3^{\circ}$ of the North Ecliptic cap) using
the Cambridge Low Frequency Synthesis Telescope at 151 MHz but did not
find any new halo sources.

Recent X-ray selected surveys of halos as well as observations aimed
at detecting the SZ effect have found many more halo candidates
suggesting that halos may not be as rare as they were once thought to
be.

Moffet and Birkinshaw (1989) first suggested that there may be a
correlation between the presence of an SZ effect and a radio halo
source, since the only three clusters A2218, A665, and CL0016+16 which
had an SZ effect detected at the time also had extended diffuse radio
emission. One of the strongest radio halos was found in A2163 in an
attempt to detect the SZ effect (Herbig \& Birkinshaw 1994). Among the
seven clusters observed at the ATCA for the SZ effect, two show
clear evidence of a radio halo (A2163, 1E0657$-$56), while
another three show faint extended emission which may be either the
result of the blending of discrete radio sources or a faint halo
(Liang 1995). It is perhaps not surprising that searches for the SZ
effect have been good at finding halo sources: the SZ effect is also
cluster-wide, thus diffuse and extended like the halo sources. Any
observation designed to search for the SZ effect will optimise the
brightness sensitivity and thus favour the detection of halos. If, in
addition, there is a physical mechanism that associates hot, luminous,
X-ray emitting atmospheres and radio halos, searches for SZ effects
which target such clusters would be expected to find radio halos
frequently.

Giovannini et al. (1999a), in their correlation of NVSS images with
the catalogue of X-ray Brightest Abell Clusters (XBAC; Ebeling et
al. 1996), found 13 candidates for diffuse radio halos. They noticed a
significant increase in the percentage of diffuse radio sources in
high luminosity clusters compared with low luminosity clusters:
27--44\% in clusters with L$_{x}>10^{45}$ erg/s as compared with 6-9\%
for L$_{x}<10^{45}$ erg/s.

We conclude that radio halos are not intrinsically rare, and appeared
to be rare from the results of early surveys partly because of the
difficulty of detecting such low surface brightness objects and partly
because of the selection criteria.

\subsection{The Link between Thermal and Relativistic Electrons}
While Giovannini et al. (1999a) found more halos in high than low
X-ray luminosity clusters, they did not find a correlation between
their radio power and the cluster X-ray luminosity. 
Here we plot the rest frame 1.4\,GHz radio power ($P_{1.4}$) against
the cluster X-ray luminosity ($L_{x}$) for only well-confirmed radio
halos (not relic sources) using the best radio data available for each
halo. Fig.~\ref{haloll} shows that there is a correlation between
radio and X-ray luminosities for
$L_{(0.1-2.4)keV}>10^{45}$\,ergs\,s$^{-1}$ clusters contrary to
Giovannini et al. (1999a) where they plotted all candidate halos using
the radio power obtained from the NVSS for each halo. Instead of
plotting radio power against X-ray luminosity, we examine the
relationship between halo radio power and cluster X-ray
temperature. Fig.~\ref{halopt} shows the 1.4 GHz integrated radio
power of cluster halos plotted against the cluster temperature,
demonstrating a steep correlation.  Since only well-confirmed radio
halos are plotted, the sample of clusters shown is by no means
complete. The apparent rareness of halos can be explained by the
steepness of the relationship shown in Fig.~\ref{halopt}: only
clusters with a high X-ray temperature at moderate redshifts are
easily detectable. The surface brightness of halos decreases with
increasing redshift at least as fast as $(1+z)^{5}$ when taking
account of the K-correction, thus the halo surface brightness rapidly
diminishes with increasing redshift. On the other hand, halos at low
redshift are also difficult to detect since they tend to be resolved
out in simple interferometric maps (or single dish observations
without a large beam-throw).

\placefigure{haloll} \placefigure{halopt}

In the 3 well-imaged cluster halos (Coma, A2163 \& 1E0657$-$56), the
extent and shape of the radio halo follows closely that of the cluster
X-ray emission (Fig.~\ref{hrir}; Deiss et al. 1997; Herbig \&
Birkinshaw in preparation).  Both the correlation shown in
Fig.~\ref{haloll}~\&~\ref{halopt} and the similarities between the
radio and X-ray morphology indicate a direct connection between the
thermal particles and the relativistic electrons responsible for the
radio emission.

\subsection{Formation of Radio Halos \label{s53}}
We favour a model for radio halos where thermal electrons in the ICM
provide the seed particles for acceleration to the ultra-relativistic
energies necessary for synchrotron radiation (Liang 1999, Brunetti et
al. 1999 \& Schlickeiser et al. 1987). This differs from the early
theories (e.g. Jaffe 1977) where the seed electrons diffused out of
radio galaxies. In the past, the possibility of accelerating thermal
electrons to relativistic energies has been dismissed, usually with
little justification.  We will examine this issue by looking at the
arguments put forward against the possibility of accelerating thermal
electrons to relativistic energies. One of the simple arguments was
that such a process would be present in every cluster and thus fail to
explain the perceived rarity of halos (Giovannini et al. 1993). As
discussed earlier, such an argument does not necessarily hold: the
strength of halos appears to be strongly related to the temperature of
the thermal gas, and halos may be in every cluster and be detectable
or not according to their brightness.  Since high temperature clusters
are relatively uncommon, so are powerful halos, but high-temperature
clusters are likely to contain halos.

A second argument against the ``thermal pool'' origin of the
ultra-relativistic electrons was that stochastic processes such as
Alfv\'en, turbulent and shock acceleration are only efficient in
accelerating electrons that are already mildly relativistic
(e.g. Eilek \& Hughes 1991 and references therein). This means that
either the seed electrons are mildly relativistic already, or an
injection process is necessary to create a substantial suprathermal
tail in the electron energy distribution. To circumvent the injection
process (not yet understood), it is attractive to invoke models that
use the already-relativistic electrons from radio galaxies as seed
particles to be re-accelerated.

However, observationally, it has also been shown that while halo
candidates were found in nearly 30\% of the high X-ray luminosity
clusters in a survey for radio halo sources, none was found in the 11
clusters selected by the existence of at least one tailed radio source
within 0.1 Abell radius (Giovannini et al. 1999b). This is further
evidence that the relativistic electrons are more likely to have
originated from the thermal pool of electrons than the radio galaxies:
the appearance of halos is more closely linked to thermal X-ray
emission than the presence of tailed radio galaxies.

A closer examination of the argument for an injection process show
that it may not always be necessary in a cluster
environment. According to Eilek \& Hughes (1991), cyclotron resonance
is responsible for the acceleration of electrons by Alfv\'en waves, and
higher energy particles resonate with lower frequency waves. Since
Alfv\'en waves have frequencies less than the ion gyrofrequency
($\Omega_{p}=eB/(m_{p}c)$), electrons need a minimum energy to be
accelerated by Alfv\'en waves. This threshold energy can be calculated
by substituting the ion gyrofrequency $\Omega_{p}$ and the dispersion
relation into the equation for cyclotron resonance:
$\omega-k_{||}v_{||}+\Omega_{e}/\gamma=0$, where $\omega$ is the wave
frequency, $k_{||}=k\cos{\phi}$ and $v_{||}=v\cos{\theta}$ are the
components of the wave-number and particle velocity parallel to the
magnetic field, and $\Omega_{e}/\gamma$ is the relativistic electron
gyrofrequency. In the low-amplitude limit the waves are
non-compressive, and the dispersion relation is given by (Krall \&
Trivelpiece 1973)
\begin{equation}
\frac{\omega^{2}}{k^{2}}=\frac{v_{A}^{2}\cos^{2}{\phi}}{1+v_{A}^{2}/c^{2}},
\end{equation}
where $v_{A}=B/\sqrt{4\pi m_{p} n_{e}}$ is the Alfv\'en speed.
Hence the minimum energy or Lorentz factor required for an electron to be accelerated by Alf\'ven waves is given by
\begin{equation}
\gamma_{min}=\frac{m_{p}}{m_{e}}(\frac{v_{A}}{\eta c})^{2}+\frac{v_{A}}{\eta c}\sqrt{(\frac{m_{p}}{m_{e}})^{2}+(\frac{\eta c}{v_{A}})^{2}},
\end{equation}
where $\eta=\cos{\theta}$.  This threshold electron energy is already
mildly relativistic ($>100$\,keV) in the limit of $(\eta c/v_{A})^{2}<
(m_{p}/m_{e})^{2}$, i.e. $n_{e} < (2\times 10^{-4}/\eta^{2}) (B/\mu
G)^{2}$\,cm$^{-3}$.  However, in a high density environment,
$\gamma_{min}$ is close to one and thermal electrons can be
accelerated by Alfv\'en waves. The environments of radio galaxies,
including clusters, were considered to be of low density since the
magnetic field strength $B$ was thought to be a few $\mu$G, which
means the density threshold is higher than the densities of most clusters.

Recent hard X-ray results from Beppo-SAX and Rossi-RXTE for Coma and
other clusters have shown the magnetic field to be $B\sim 0.2\mu$G if
the excess hard X-ray emission is due to inverse Compton scattering of
relativistic electrons by the CMB (e.g. Fusco-Femiano et al. 1999;
Rephaeli et al. 1999; Valinia et al. 1999). Thus the density threshold
is now $\sim 8\times 10^{-6}/\eta^{2}$\,cm$^{-3}$, which makes most
parts of clusters high density environments. For example, in the
centres of clusters where $n_{e}\sim 10^{-3}$, the minimum energy
required is just a few keV, and in the outer parts of a cluster where
$n_{e} \sim 10^{-4}$ the minimum energy is still just a few
tens of keV. Electrons with energies of a few tens of keV are readily
available in intracluster plasma of hot clusters. Therefore, it is
possible to accelerate thermal electrons in clusters through resonance
with Alfv\'en waves. Further, Dogiel (1999) has shown that it is
possible to produce a substantial suprathermal tail in the electron
energy distribution through second order Fermi acceleration in cluster
environments, so that additional seed electrons are naturally present
in clusters without involving injection from radio galaxies. 

On the one hand, the higher the X-ray luminosity of a cluster, the
higher the density of thermal electrons; and on the other hand, the
higher the X-ray temperature of a cluster, the higher the fraction of
high energy electrons. These two effects multiply to increase the
number of electrons above a threshold energy for efficient
acceleration processes. 

Both merging activity and the electron temperature may be responsible
for the production of radio halos. A possible scenario would be that
the initial merging activity provides the energy for accelerating
electrons from the suprathermal tail of the energy distribution (where
the hotter clusters have more power) to ultra-relativistic
energies. Since cluster magnetic field strengths are less than
$3$\,$\mu$G, the dominant energy loss mechanism for relativisitic
electrons is inverse Compton scattering of the cosmic microwave
background radiation. Thus the typical lifetime of an electron that
emits at 1.4\,GHz is $t_{age}\sim 8\times 10^{7} \sqrt{B/(1\mu{\rm
G})} (1+z)^{-9/2}$\,yr. If we assume a magnetic field of $B\sim
0.2\,\mu$G, then the lifetime of an electron that emits at 1.4\,GHz in
1E0657$-$56 is $\sim 10^{7}$\,yr. After the shocks have disappeared,
radio halos like that of 1E0657$-$56 may be maintained by in-situ
electron acceleration in the residual turbulence. In the case of
1E0657$-$56, we found (Fig.~\ref{hrir}) that the radio halo emission
is enhanced at the peak of the X-ray emission as well as that of the
galaxy distribution. The enhancement of radio emission at the peak of
the X-ray emission is naturally explained by our model where we expect
the highest density of relativistic electrons at the density peak of
the thermal electrons. We also expect that the galaxies streaming
through the hot intracluster gas would maintain local turbulence and
hence inject energy into the particles, producing enhanced radio
emissions at the galaxy concentrations. Deiss \& Just (1996) found
through their calculations that it is possible to have turbulent
velocities of several hundred km\,s$^{-1}$, which could considerably
enhance the stochastic acceleration rates.

Cooling flow clusters are thought to be relaxed and devoid of merging
activity. Most of the clusters shown in Fig.~\ref{halopt} are
non-cooling flow clusters.  This does not imply that mergers are the
{\it critical} element in radio halo formation, since selection
effects act to remove clusters from the sample in Fig.~\ref{halopt}:
cooling flow clusters are more likely to have significant central
radio sources than non-cooling flow clusters (e.g. Peres et
al. 1998). To our knowledge, no cluster with a strong cooling flow has
been observed with sufficient dynamic range and surface brightness
sensitivity to test whether or not it follows the $P_{1.4}-kT_{x}$
trend shown in Fig.~\ref{halopt}. To illustrate the need for proper
observations with high brightness sensitivities, we give as an
example, RXJ1347-11, a strong cooling flow with the high gas
temperature of $\sim 12.5$\,keV (Allen \& Fabian 1998) which has been
observed by the NVSS with no obvious detection. However, the NVSS does
not have enough brightness sensitivity to detect, in RXJ1347-11, a
halo similar to that in 1E0657$-$56 because of the high redshift
($z\sim 0.45$) of the cluster (expected signal of $\sim 1.3$\,mJy per
$45^{''}$ beam) and the high noise levels in the NVSS image ($\sim
0.5$\,mJy per $45^{''}$ beam).

\section{CONCLUSIONS}
We have found a powerful radio halo in the cluster 1E0657$-$56. At a
rest frame 1.4\,GHz radio power of $(4.3\pm0.3)\times
10^{25}$\,W\,Hz$^{-1}$, it is one of the most powerful radio halo
sources. It has a steep spectral index of $\alpha_{4864}^{1344} \sim
-1.2$ typical of known halos. The brightness distributions of the
radio halo and X-rays from the cluster gas are remarkably similar,
suggesting a direct relationship between the ultra-relativistic
electrons responsible for the synchrotron emission and the thermal
intracluster gas. As further evidence for the radio/X-ray connection,
we have found a steep correlation between the radio power of the halo
and the X-ray temperature of the intracluster gas ($P_{1.4}-kT_{x}$)
from the 10 confirmed cluster radio halos.  We favour an explanation
for the origin of radio halo sources, where the high energy tail of
the thermal electron distribution is boosted to ultra-relativistic
energies, thus providing a natural link between the halo radio power
and X-ray gas temperature. Detailed calculations for such a model will
be given in a future paper.  Finally, it is important for our
understanding of the origin of radio halo sources to establish the
robustness of the $P_{1.4}-kT_{x}$ correlation by observing a
temperature selected sample of clusters, and to test the mechanism by
searching for halos in clusters with strong cooling flows but high
temperature.

In addition, we have re-analysed the X-ray spectroscopic data using
both ASCA GIS and ROSAT PSPC data for 1E0657$-$56 and found the best fit
temperature to be $kT_{x}=14.5^{+2.0}_{-1.7}$\,keV consistent with it
being one of the hottest known clusters, as claimed by Tucker et
al. (1998). The use of the PSPC data enabled us to determine the
soft X-ray absorption to a better accuracy than previous results using
ASCA data alone. We found the best fit neutral hydrogen column density
to be consistent with the Galactic value given by radio-astronomical
surveys, contrary to the much higher column density claimed by Yaqoob
(1999) using ASCA data alone.

\acknowledgments

We are grateful to Sarah Maddison for help with the ATCA observations;
Emilio Falco, Massimo Ramella and Wallace Tucker for providing the NTT
image prior to publication; Christian Br\"uns and Lister
Staveley-Smith for taking a H {\small I} measurement with Parkes Radio
Telescope towards the cluster; and E. M. Arnal for providing 21cm
measurements of the neutral Hydrogen column density from the IAR high
sensitivity H {\small I} survey. H.L. would like to thank Ron Ekers
for encouragement and helpful discussions, the ATNF for hospitality.
The MOST is operated with the support of the Australian Research
Council and the Science Foundation for Physics within the University
of Sydney.  The Australia Telescope is funded by the Commonwealth of
Australia for operation as a National Facility managed by CSIRO. We
would like to acknowledge the use of data from the HEASARC online
service and the ROSAT public archive, and the use of the {\small
Karma} package (\url{http://www.atnf.csiro.au/karma}) for the
overlays.

\clearpage

\figcaption[xrayspec.ps]{X-ray spectra from the ASCA GIS and ROSAT
PSPC. The observed spectra are shown with error bars. A Raymond-Smith
thermal spectrum with $kT_{x}\sim 14.5$\,keV, $A\sim 0.33$ solar and
absorption $N_{H} \sim 4.2\times 10^{20}$\,cm$^{-2}$, convolved with
instrumental responses, is shown as a histogram plot. A good fit to
the data is obtained, with reduced $\chi^{2}$ of
1.03/d.o.f. \label{ascasp}}

\figcaption[xrayspec2.ps]{X-ray spectra from the ASCA GIS and ROSAT
PSPC, as in Fig.~\ref{ascasp}. The superimposed model spectra, shown
as a histogram plot, arise from a Raymond-Smith thermal spectrum with
$kT_{x}\sim 11.3$\,keV, $A\sim 0.2$ solar and absorption $N_{H} \sim
15\times 10^{20}$\,cm$^{-2}$ after convolution with the instrumental
responses. This model, consistent with the results of Yaqoob (1999)
from ASCA data alone, is a poor fit, with a reduced
$\chi^{2}=1.56$. The data require less soft X-ray absorption than in
this model. \label{ascasp2}}

\figcaption[pspch50.843s60.ps,pspch50.1344s60.ps,pspch50.2350s60.ps,pspch50.4864s60.ps,pspch50.5888s60.ps,pspch50.8832s60.ps]{Low
resolution radio images overlaid on grey scale PSPC hard-band
(0.5--2.0 keV) image smoothed with a $50^{''}$ Gaussian: a) MOST
contour image at 843\,MHz smoothed to a beam size of $60^{''}$; b)-e)
ATCA contour images with a $60^{''}$ beam at 1.3, 2.4, 4.9, 5.9, and
8.8\,GHz respectively; only the shortest spacings ($<3600\lambda$) are
used. Contour levels are $(-3,3,6,12,24,48,96,192,384)\times\sigma$,
where rms noise are $\sigma\sim 1100, 51, 110, 56, 65, 56 \mu$Jy/beam
for frequencies of 0.8, 1.3, 2.4, 4.9, 5.9, and 8.8\,GHz
respectively. \label{allfreq} }

\figcaption[rxjcen.1344.1.ps,pts.1344.reg.1.ps]{a) A 1.3\,GHz ATCA
image towards 1E0657$-$56 at a resolution of $ 6.5^{''}\times
5.9^{''}$, using all the data (uniform weighting). The noise level in
the image is $44 \mu$Jy/beam. b) A 1.3\,GHz image using only the long
baseline data ($>5000\lambda$) with only the CLEAN components within
the marked region `1' restored (beam size $ 6^{''}$). The unresolved
sources are marked by a circle and the two extended sources (A and C)
are marked by the small areas that were used to integrate their total
flux. The sources are marked by letters and the three regions marked
from `1' to `3' correspond to regions used for total flux estimates.
\label{21cm}}

\figcaption[halos20.1344.hris10.1.ps,halos20.1344.nttR.ps,rxjhris10.nttR.ps]{a)
A grey scale image of the radio halo at 1.3 GHz with resolution
$24^{''}\times 22^{''}$ after the subtraction of catalogued sources
(Table~\ref{ptsrc}) is overlaid with a ROSAT HRI contour image
smoothed with a Gaussian of $10^{''}$ width.  The contour levels are
$0.35,0.45,0.6,0.8,1.0,1.2,1.4,1.6,1.8,2.0,2.2$\,HRI\,cts/s. b) The
same radio image in contours overlaid on a ESO New Technology
Telescope (NTT) R-band image (courtesy of E. Falco and M. Ramella)
with the same image scale as a). The radio contours are
$(3,6,12,18,24)\times\sigma$, where the noise $\sigma$ in the
radio image is $90\mu$Jy/beam. c) The same HRI contours overlaid on
the NTT R-band image. \label{hrir}}

\figcaption[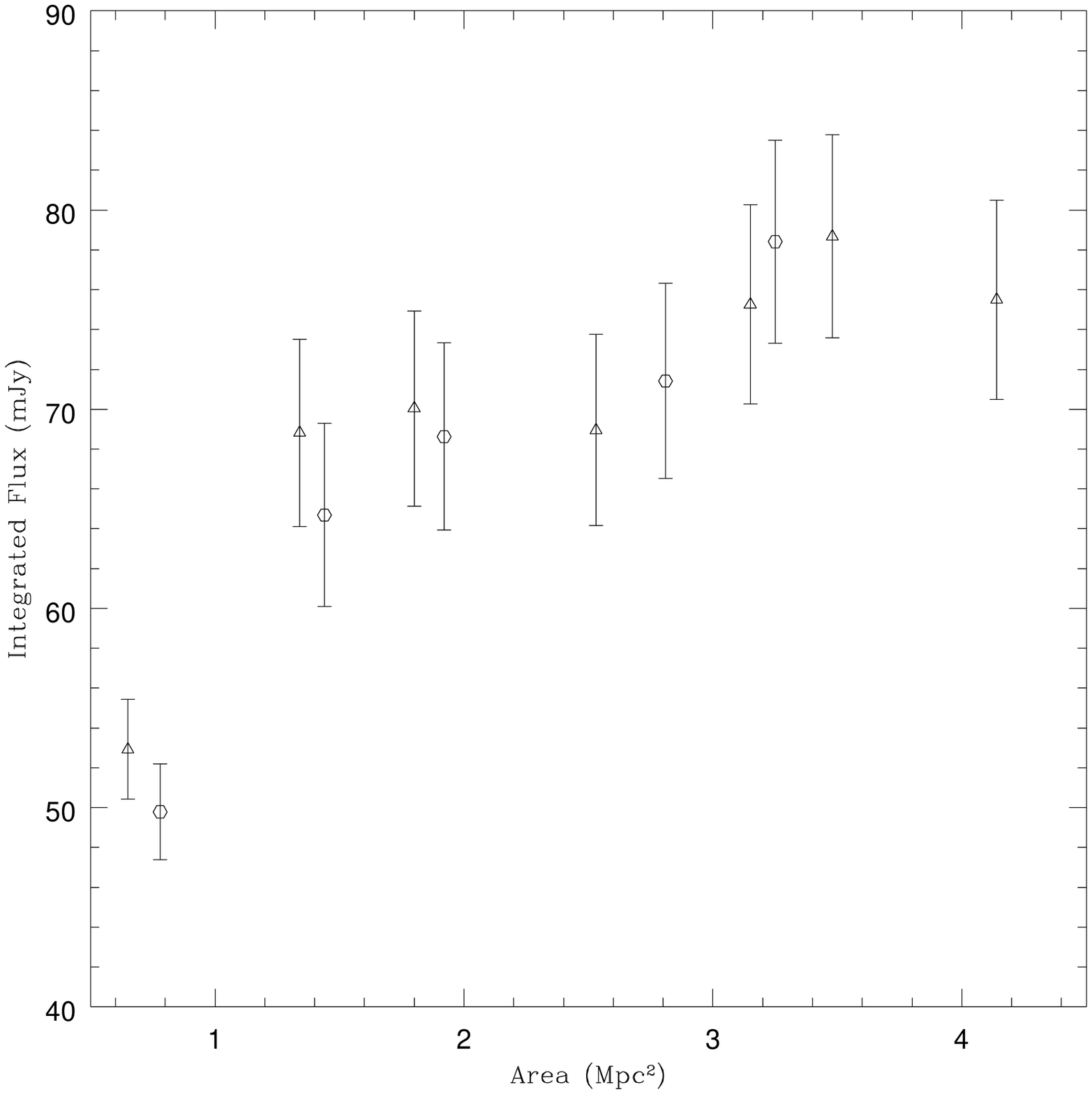]{Integrated diffuse halo flux density versus
the area of integration. The open circles are integrated flux
densities from an image made with a source-subtracted uv data-set and
smoothed to $60^{''}$ resolution using only the short baseline
($<3600\lambda$) data. The triangles are integrated flux densities
from a high resolution image shown in Fig.~\ref{21cm}a but with the
embedded sources obtained from Fig.~\ref{21cm}b subtracted afterwards.
The error bars are $1\sigma$ errors. \label{totflux}}

\figcaption[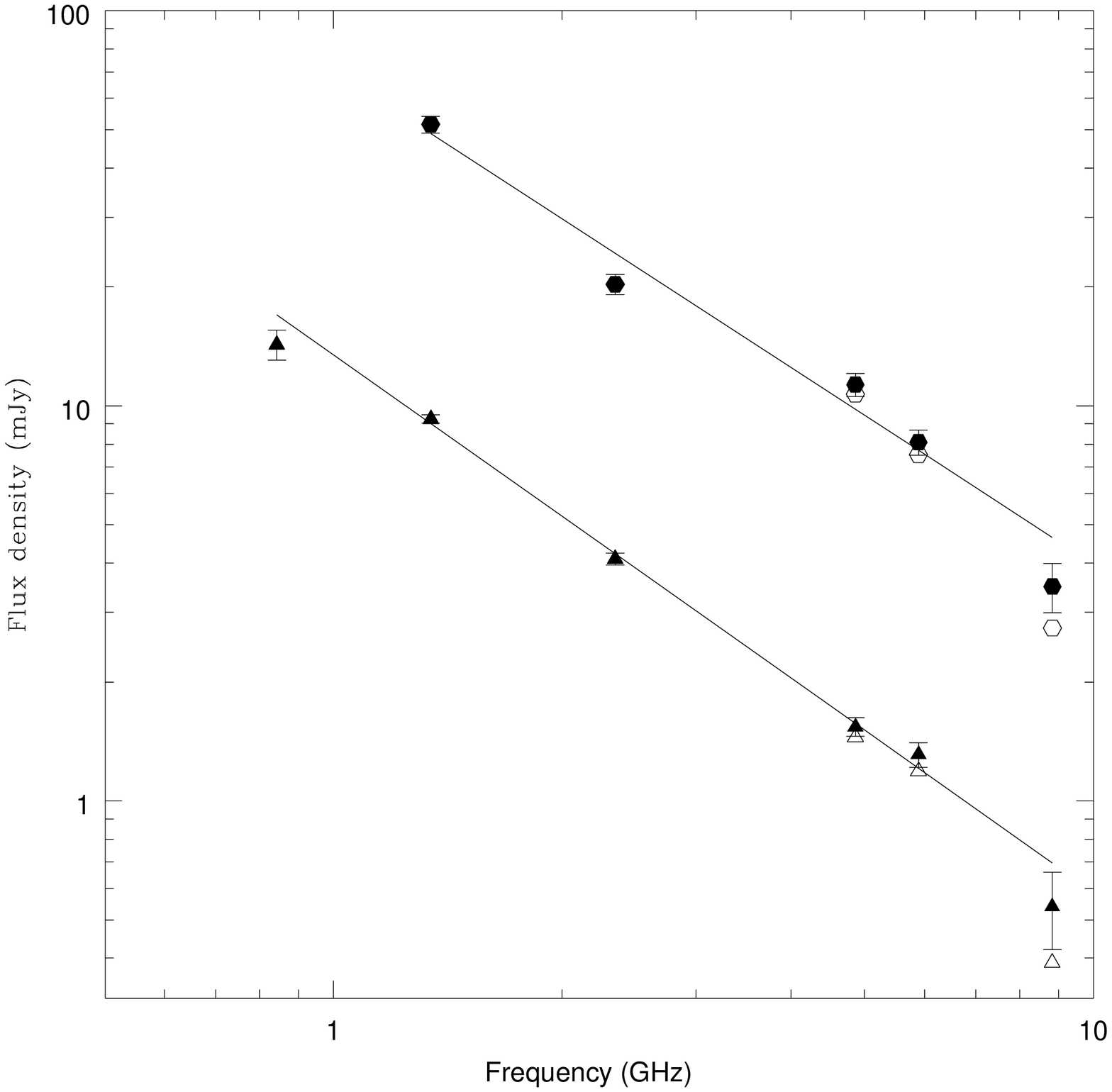]{Radio spectra from 0.843\,GHz to 8.8\,GHz of
the cluster halo in 1E0657$-$56. The spectrum of the central region
(region 3 in Fig.~\ref{21cm}(b)) is marked by triangles. For the
larger region (region 2 in Fig.~\ref{21cm}(b)) the spectrum is marked
by circles.  The filled data points have been corrected for the SZ
effect. The uncorrected flux are represented as open symbols. The
error bars are $1\sigma$ errors. The straight lines are the best least
square fits to a power law spectrum.
\label{halosp}}

\figcaption[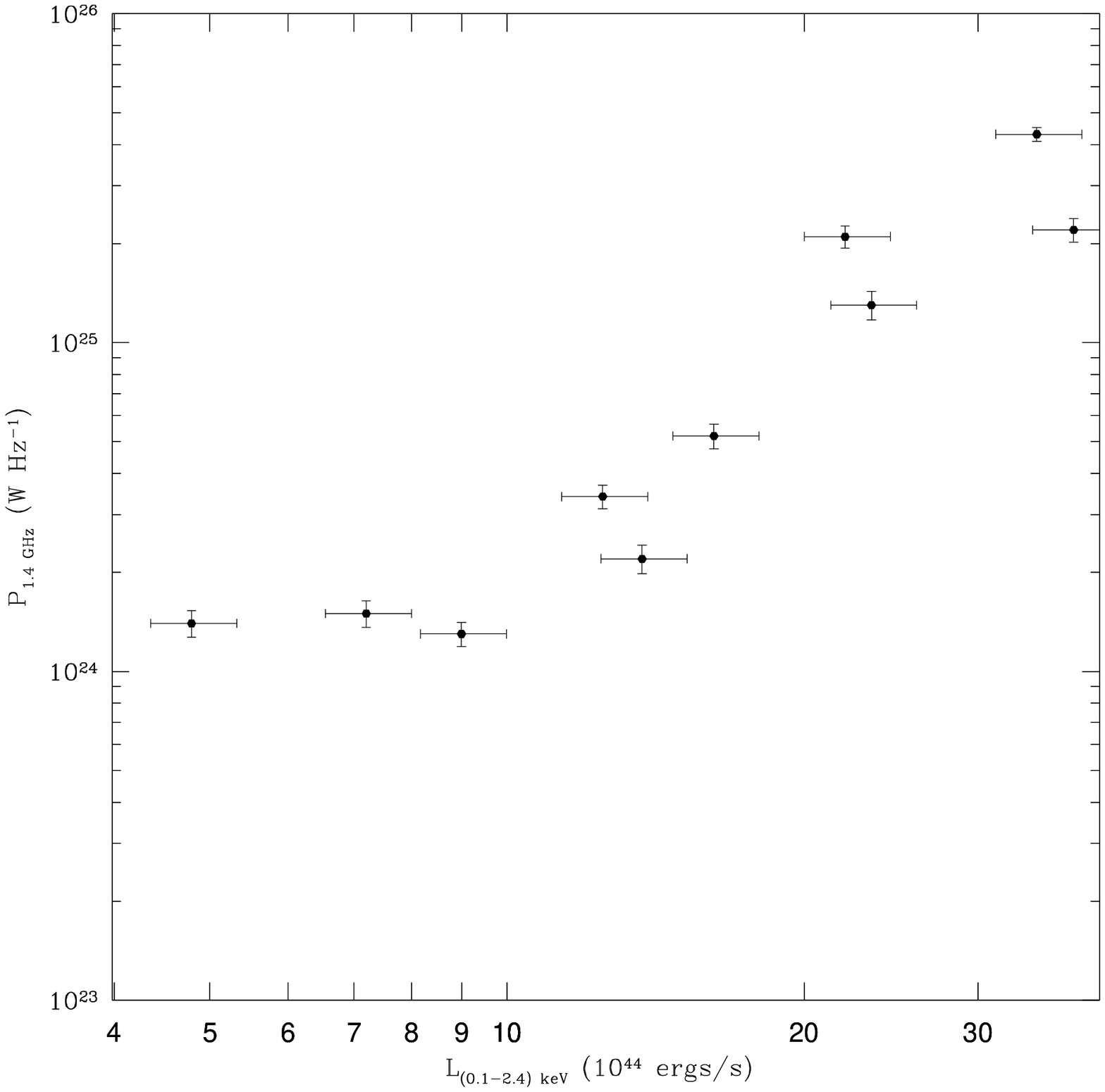]{Radio power $P_{1.4}$ at 1.4 GHz (rest frame)
versus the cluster X-ray luminosity (0.1-2.4)\,keV for all the
confirmed radio halo sources. Formal errors of $\sim 1\sigma$ are
given for the halo power where it is known, in cases where the errors
are not obviously stated in the literature, a 10\% error is assumed. A
10\% error is assumed for the X-ray luminosity. X-ray luminosities are
obtained from Ebeling et al. 1996 and this paper. The radio powers are
from Herbig \& Birkinshaw in preparation, Reid et al. 1999, Giovannini
et al. 1993, Feretti et al 1997a, 1997b, Giovannini et al. 1999a,
Liang et al. in preparation and this paper. \label{haloll}}

\figcaption[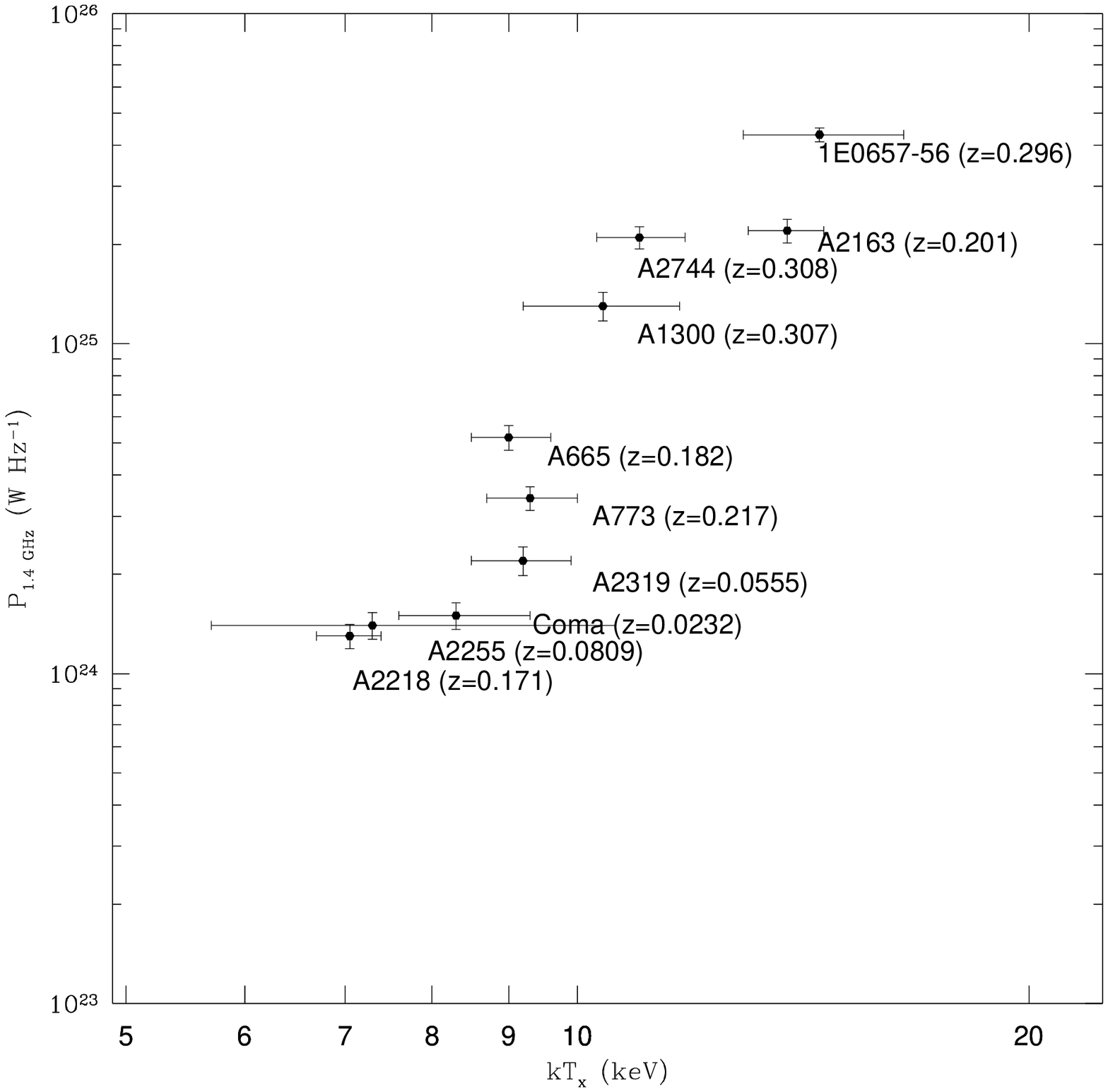]{Radio power $P_{1.4}$ at 1.4 GHz (rest frame)
versus the cluster X-ray temperature $kT_{x}$ for all the confirmed
radio halo sources. The error bars for the intracluster gas
temperature correspond to a 90\% confidence limit. Formal errors of
$\sim 1\sigma$ are given for the halo power where it is known, in
cases where the errors are not obviously stated in the literature, a
10\% error is assumed.  The X-ray data are obtained from Allen \&
Fabian 1998, Markevitch et al. 1998, Pierre et al. 1999, Mushotzky \&
Scharf 1997, David et al. 1993 and this paper. The radio data are the same as in Fig.~\ref{haloll}. \label{halopt}}

\begin{deluxetable}{cccccccccr}
\tablecaption{Summary of Radio Observations \label{t:obs}}
\tablecolumns{10}
\tablewidth{0pt}
\tablehead{
\colhead{Date} & \colhead{Frequency}   & \colhead{Config.}   &
\multicolumn{6}{c}{Centre (J2000)} & \colhead{$t_{obs}$} \\
\colhead{} & \colhead{(MHz)} & \colhead{} & \colhead{h} & \colhead{m} & \colhead{s} & \colhead{$\circ$} & \colhead{$'$} & \colhead{$''$} & \colhead{(hr)}}
\startdata
Dec 1996 & 8768/8896 & 210 & 06 & 58 & 32.7 & $-$55 & 57 & 19 & 56.6 \\
         & 4800/4928 & 210 & 06 & 58 & 32.7 & $-$55 & 57 & 19 & 11.2 \\
         & 5824/5952 & 210 & 06 & 58 & 32.7 & $-$55 & 57 & 19 & 11.4 \\
May 1997 & 1344/2240 & 6B & 06 & 58 & 32.7 & $-$55 & 57 & 19 & 11.2 \\
Jun 1997 & 1344/2240 & 750A & 06 & 58 & 32.7 & $-$55 & 57 & 19 & 10.0 \\
Jul 1998 & 1344/2496 & 750E & 06 & 58 & 32.7 & $-$55 & 57 & 19 & 10.8 \\
         & 8768/4800 & 750E & 06 & 58 & 32.7 & $-$55 & 57 & 19 & 9.7 \\
Jan 1998 & 843       & MOST & 07 & 00 & 00.0 & $-$56 & 35 & 24 & 12.0 \\
\enddata


\end{deluxetable}

\begin{deluxetable}{cccccccrrrr}
\tablecaption{Radio Flux Density of the Discrete Sources\label{ptsrc}}
\tablecolumns{11}
\tablewidth{0pt}
\tablehead{
\colhead{Source}  & \multicolumn{6}{c}{Peak Position (J2000)} & \multicolumn{4}{c}{Flux Density (mJy)} \\
\colhead{} & \colhead{h} & \colhead{m} & \colhead{s} & \colhead{$\circ$} & \colhead{$'$} & \colhead{$''$} & \colhead{$S_{1344}$} & \colhead{$S_{2350}$} & \colhead{$S_{4864}$} & \colhead{$S_{8832}$}}
\startdata
A & 06 & 58 & 37.9 & $-$55 & 57 & 25  & $19.1\pm 1$    & $11.8\pm 0.6$  & $6.3\pm 0.3$ & $3.3\pm 0.2$ \\
B & 06 & 58 & 34.1 & $-$55 & 57 & 54  & $0.5\pm 0.05$  & $<0.3$         & $<0.3$       & $<0.4$   \\
C & 06 & 58 & 42.2 & $-$55 & 58 & 37  & $37\pm 2$  & $21\pm 2$    & $8.4\pm 0.5$ & $3.2\pm 0.5$ \\
D & 06 & 58 & 23.4 & $-$55 & 56 & 41 & $1.1\pm 0.06$  & $0.4\pm 0.07$  & $<0.3$       & $<0.4$    \\
E & 06 & 58 & 27.2 & $-$55 & 56 & 08 & $0.4\pm 0.05$  & $<0.3$         & $<0.3$       & $<0.4$    \\
F & 06 & 58 & 24.3 & $-$55 & 55 & 13 & $0.6\pm 0.05$  & $0.4\pm 0.07$  & $<0.3$       & $<0.4$    \\
G & 06 & 58 & 19.3 & $-$55 & 58 & 43 & $0.8\pm 0.05$  & $0.4\pm 0.07$  & $<0.3$       & $<0.4$    \\
H & 06 & 58 & 16.6 & $-$55 & 58 & 23 & $0.4\pm 0.05$  & $<0.3$         & $<0.3$        & $<0.4$    \\
\enddata

\tablecomments{The upper limits in flux densities are $3\sigma$, but the errors quoted are $1\sigma$.}

\end{deluxetable}

\plotone{xrayspec.ps}

\plotone{xrayspec2.ps}




\plotone{totflux2.ps}

\plotone{halospf.ps}

\plotone{haloll.ps}

\plotone{halopt.ps}

\end{document}